\documentclass[%
reprint, 
amsmath,amssymb,
aps,
prd,
]{revtex4-2}

\usepackage{graphicx}
\usepackage{ctable}        
\usepackage{array}         
\usepackage{xcolor}        
  \definecolor{navy}{HTML}{2c1fa5}
\usepackage[final]{hyperref} 
\hypersetup{
	colorlinks=true,         
  linkcolor=navy,          
	citecolor=navy,          
	filecolor=magenta,       
	urlcolor=navy         
}

\begin{document}

\title{Dynamical evolution of {$U(1)$} gauged Q-balls in axisymmetry}

\author{Michael P. Kinach}
 \email{mikin@physics.ubc.ca}
\author{Matthew W. Choptuik}%
 \email{choptuik@physics.ubc.ca}
\affiliation{%
  Department of Physics and Astronomy, University of British Columbia,\\
  6224 Agricultural Road, Vancouver, British Columbia, V6T 1Z1, Canada
}%

\date{\today}

\begin{abstract}
  We study the dynamics of $U(1)$ gauged Q-balls using fully
  non-linear numerical evolutions in axisymmetry. Focusing on two models
  with logarithmic and polynomial scalar field potentials, we
  numerically evolve perturbed gauged Q-ball configurations in order to
  assess their stability and determine the fate of unstable
  configurations. Our simulations suggest that there exist both stable
  and unstable branches of solutions with respect to axisymmetric
  perturbations. For solutions belonging to the stable branch, the
  gauged Q-balls respond to the perturbations by oscillating
  continuously or weakly radiating before returning to the initial
  configuration. 
  For the unstable branch, the solutions are eventually
  destroyed and can evolve in several ways, such as dispersal of the
  fields to infinity or fragmentation into smaller gauged Q-balls. In
  some cases, we observe the formation of ring-like structures which we
  call ``gauged Q-rings". 
  We also investigate the stability of gauged
  Q-balls when the gauge coupling is small, finding that the behaviour
  of these configurations closely resembles that of ordinary
  (non-gauged) Q-balls.
\end{abstract}

\maketitle


\section{\label{sec:intro} Introduction }

Solitons are a fundamental prediction of many physical theories. They
are characterized as stable, localized solutions to non-linear field
equations that behave in many ways like particles. Broadly speaking,
solitons can be classified as either topological or non-topological.
Topological solitons owe their existence and stability to the specific
topological constraints of a given model. 
Non-topological solitons, in contrast, can arise simply due to the
balance of attractive and repulsive self-interactions in the theory. In
addition, the stability of non-topological solitons is often associated
with the presence of conserved charges which emerge from the theory's
underlying symmetries (though one can also construct solitonic
configurations in the absence of such charges \cite{Seidel1991}).

Perhaps the simplest example of a non-topological soliton in field theory is
the Q-ball: a stable, localized solution of a complex scalar field theory with
a non-linear attractive potential and a global or gauge $U(1)$ symmetry.  In
recent years, Q-balls have attracted significant attention due to their
prevalence in supersymmetric theories \cite{Kusenko1997} and their possible
cosmological consequences.  In particular, it has been suggested that Q-balls
may be relevant for baryogenesis \cite{Kasuya2000,Dine2003}, cosmological phase
transitions \cite{Frieman1988,Kusenko1997b}, and the dark matter problem
\cite{Kusenko1998,Kusenko2001}. The formation of Q-balls could also lead to
detectable gravitational wave signatures \cite{Kusenko2008}. 
However,
regardless of their physical applications, Q-balls are also interesting from a
theoretical perspective as stable, particle-like objects that can be
constructed from smooth classical fields
and that have vanishing topological charge.

The properties of Q-balls under a global $U(1)$ symmetry have been studied
extensively in the literature. Starting with the work of Rosen
\cite{Rosen1968}, Q-ball solutions have been found in a variety of
physically-motivated models using various scalar field potentials (see
\cite{Nugaev2020} for a recent review).  For some special potentials, the
equations can be solved exactly
\cite{Rosen1969,Theodorakis2000,Gulamov2013,Bazeia2016}, but in the general
case one must use approximations or numerical methods in order to determine the
characteristic features of Q-balls. Associated with each solution in a given
model, there is a conserved energy $E$ and a conserved Noether charge $Q$ (from
which the Q-ball gets its name) corresponding to the particle number.  Each
solution is also characterized by an internal oscillation frequency $\omega$
which can be interpreted as the chemical potential of the configuration
\cite{Heeck2021c}.  In addition to ordinary (ground state) Q-balls, one can
construct excited Q-balls which have additional radial nodes or non-zero
angular momentum \cite{Friedberg1976,Volkov2002,Kleihaus2005,Mai2012,Almumin2022}. The basic
theory has also been extended by coupling Q-balls to gravity
\cite{Lynn1989,Multamaki2002,Tamaki2011}, by introducing a massless or massive
gauge field \cite{Rosen1968b,Lee1989,Heeck2021d}, and by considering
non-spherical configurations such as Q-tubes \cite{Sakai2011}, Q-rings
\cite{Axenides2001}, and composite systems of Q-balls \cite{Copeland2014}.

When the global $U(1)$ symmetry of the theory is gauged, the Q-balls acquire an
electric charge and are known as \textit{gauged Q-balls} \cite{Lee1989}. Gauged
Q-balls have properties that can differ significantly 
when contrasted to their global
(non-gauged) counterparts. The presence of a massless gauge field introduces a
long-range repulsive force that can destabilize the solutions for large
gauge couplings. This repulsive force can give rise to novel scalar field
configurations such as Q-shells \cite{Arodz2009,Tamaki2014,Heeck2021b}, but it
can also place limits on the maximum size and charge of gauged Q-balls for some
scalar field potentials \cite{Tamaki2014,Gulamov2015}. The existence of this
maximal charge corresponds with the limits of the allowed range of the
frequency $\omega$, and in general the gauged Q-ball configurations cannot be
uniquely characterized by the value of $\omega$ \cite{Gulamov2015}. 
Despite these differences, there exists a correspondence which allows
for the properties of gauged Q-balls to be approximated from the
properties of non-gauged (global) Q-balls, which are often much simpler
\cite{Heeck2021}.
In addition, when the interaction between the scalar field and gauge field is
weak, gauged Q-balls are expected to closely resemble their non-gauged
counterparts \cite{Gulamov2014}.

One of the essential properties of Q-balls relates to their dynamical
stability. 
In order to be physically viable, solitons must be robust against
perturbations. However, the problem of establishing
the stability of solitons is often complicated by the non-linear nature of
the governing field equations. In some cases, linear perturbation analyses and
stability theorems can be applied to determine the expected regions of
stability and instability.

For non-gauged Q-balls, it has been shown that
the simple relation
\begin{equation} \label{eqn:VK-criteria}
  \frac{\omega}{Q} \frac{dQ}{d\omega} < 0
\end{equation}
serves as an effective criterion for establishing regions of stability
\cite{Friedberg1976,Correia2001}.  However, the case of gauged Q-balls is
more complicated due to the presence of the repulsive gauge field. It was
recently shown in Ref.~\cite{Panin2017} that the sign of $dQ/d\omega$ cannot be used
to assess the stability of gauged Q-balls in the general case. In the absence
of such a criterion, one can still analyze the stability of these solutions
using (among other alternatives) a numerical approach: dynamically evolving
perturbed configurations through direct solution of the non-linear equations of
motion.  This method was applied in Ref.~\cite{Panin2017} to show that gauged
Q-balls in several models can be stable with respect to spherical perturbations.
However, it remains an open question as to whether gauged Q-balls can
be classically stable against decay from more general perturbations beyond
spherical symmetry.  In addition, the instability mode for non-gauged Q-balls
is always spherical \cite{Smolyakov2018}, but it is not known whether
gauged Q-ball decay can be mediated by non-spherical modes.

In this paper, we make progress toward understanding some aspects of gauged
Q-ball dynamics by performing fully non-linear numerical simulations of the
field equations in axisymmetry.  There are two main questions we shall
explore:
(i) what is the range of stability of gauged Q-balls in axisymmetry? And (ii),
what is the final fate of those configurations which are unstable? To answer
these questions, we construct spherical gauged Q-ball initial data 
using a numerical shooting technique.
We then assess the stability of these configurations by dynamically
perturbing the system and observing the subsequent behaviour.

Numerical results presented below suggest that there exist both stable and
unstable branches of solutions in axisymmetry. We find that stable gauged
Q-balls, when perturbed, can survive over timescales which are long
compared to the dynamical time with no evidence of measurable growing
modes which destroy the configuration. These solutions respond to
perturbation by oscillating continuously or weakly radiating before
returning to the initial configuration. Unstable gauged Q-balls, in
contrast, are typically short-lived and can decay in one of several
ways. Some unstable solutions break apart into many smaller gauged
Q-balls or shed scalar field until they relax into a smaller stable
configuration. Other unstable solutions fragment into non-spherical
ring-like structures which propagate away from the axis of symmetry
and can survive for some time. 
In addition, for the case of a logarithmic potential we observe that the
maximum magnitude of the scalar field can grow without bound.  We
interpret this behaviour as a consequence of the potential being
unbounded from below. Finally, we test the effect of the gauge coupling
strength on the stability, finding that gauged Q-balls closely resemble
their non-gauged counterparts when the coupling is small.

This paper is organized as follows: in Sec.~\ref{sec:eom}, we present the
equations of motion of the theory. In Sec.~\ref{sec:initdata}, we discuss
the procedure for obtaining axisymmetric initial data which is used in the
numerical evolutions. In Sec.~\ref{sec:diagnos}, we briefly discuss the
types of perturbations that are applied to the system.  In
Sec.~\ref{sec:results}, we present the results for several representative
evolutions.  We conclude with some final remarks in
Sec.~\ref{sec:conclusion}.

Throughout this work, we use natural units where $c=\hbar=1$ and employ the
metric signature $(-,+,+,+)$. We focus on unexcited gauged Q-ball solutions
(those for which the scalar field modulus attains only one maximum).  For
brevity, we will use the term ``Q-ball" interchangeably with ``gauged Q-ball"
when the distinction between the gauged and non-gauged solutions is 
obvious by context.

\section{\label{sec:eom} Equations of Motion}

The Lagrangian density of the theory takes the form
\begin{equation}
  \mathcal{L} = -\left(D_\mu\phi\right)^* D^\mu\phi-V\left(|\phi|\right)-\frac{1}{4}F_{\mu\nu}F^{\mu\nu},
  \label{eqn:gauged-lagr}
\end{equation}
where $\phi$ is the complex scalar field, $F_{\mu\nu}=\partial_{\mu} A_\nu -
\partial_{\nu} A_\mu$ is the electromagnetic field tensor for the $U(1)$ gauge
field $A_\mu$, and $D_\mu = \nabla_\mu-ieA_\mu$ denotes the gauge covariant
derivative with coupling constant $e$. Here, $V(|\phi|)$ is a $U(1)$-invariant
scalar field potential that permits Q-ball solutions in the limit $e\rightarrow
0$. In this work, we consider the following scalar field potentials:
\begin{align}
  V_\text{log}(|\phi|)&=-\mu^2|\phi|^2\ln(\beta^2|\phi|^2),\label{eqn:log}\\
  V_\text{6}(|\phi|)&=m^2|\phi|^2-\frac{k}{2}|\phi|^4+\frac{h}{3}|\phi|^6,\label{eqn:poly}
\end{align}
where $\mu$, $\beta$, $m$, $k$, and $h$ are assumed to be positive, real-valued
parameters of the potentials. The potential \eqref{eqn:log} has previously been
studied in various forms in Refs.~\cite{Rosen1969,Marques1976,Gulamov2014,Dzhunushaliev2013,Tamaki2014,Panin2017,Smolyakov2018}
while potential \eqref{eqn:poly} has been studied in Refs.~\cite{Lee1989,Volkov2002,Kleihaus2005,Loginov2020,Heeck2021c,Heeck2021,Almumin2022,Loiko2022}. Further details about the scalar potentials \eqref{eqn:log}
and \eqref{eqn:poly} will be discussed in the sections that follow.

The evolution equations for the theory can be found by varying the Lagrangian
density \eqref{eqn:gauged-lagr} with respect to the scalar and gauge fields to
obtain
\begin{align}
  D_\mu D^\mu \phi - \frac{\partial}{\partial \phi^*}V(|\phi|)&=0, \label{eqn:eom-a}\\
  -\nabla_\mu F^{\mu\nu}-ie\phi(D^\nu\phi)^*+(ie\phi^*)D^\nu\phi&=0. \label{eqn:eom-b}
\end{align}
From \eqref{eqn:eom-b} we identify the conserved Noether current
\begin{equation} \label{eqn:j-current}
  j^\mu=-i(\phi^* D^\mu\phi-\phi(D^\mu\phi)^*)
\end{equation}
which corresponds with invariance of the theory under the $U(1)$ transformations
\begin{align}
  \phi &\rightarrow e^{-i e \alpha(x)} \phi,\\
  A_\mu &\rightarrow A_\mu-\partial_\mu\alpha(x).
\end{align}
The conserved current \eqref{eqn:j-current} can be integrated to obtain a
conserved Noether charge $Q=\int j^{0}\,d^3x$. Also associated with the theory
is the energy-momentum tensor
\begin{equation} \label{eqn:emt}
  \begin{split}
  T_{\mu\nu}=&F_{\mu\alpha} F_{\nu\beta} g^{\beta\alpha} -\frac{1}{4}g_{\mu\nu}F_{\alpha\beta}F^{\alpha\beta}\\
  &+D_\mu\phi(D_\nu\phi)^*+D_\nu\phi(D_\mu\phi)^*\\
  &-g_{\mu\nu}(D_\alpha\phi(D^\alpha\phi)^*+V(|\phi|))
  \end{split}
\end{equation}
and the corresponding conserved energy $E=\int T_{00}\,d^3x$.

To solve the field equations of motion, we adopt the usual cylindrical 
coordinates $(t,\rho,\varphi,z)$ and
write the spacetime line element as
\begin{equation}
  ds^2=-dt^2+d\rho^2+\rho^2d\varphi^2+dz^2.
\end{equation}
In three spatial dimensions, computational constraints would limit the range of
possible field configurations that could be explored. We therefore reduce the
computational complexity of the problem by imposing axisymmetry on the system:
no dependence of any of the fields on the azimuthal angle $\varphi$ is assumed.
This results in a system of six coupled non-linear partial differential
equations which are described in App.~\ref{apx:evo-eqns}.

Evolution of the fields is subject to the constraints
\begin{align}
  \nabla\cdot\vec{E}&=\rho_c\, , \label{eqn:divE}\\
  \nabla\cdot\vec{B}&=0\, \label{eqn:divB},
\end{align}
where $\vec{E}$ and $\vec{B}$ are the electric and magnetic fields,
respectively, and $\rho_c$ is the electric charge density
\begin{equation}
  \rho_c=ie(\phi^*\partial_t\phi-\phi\partial_t\phi^*)+2e^2A_0\phi\phi^*.
\end{equation}
Equation \eqref{eqn:divB} will be trivially satisfied in axisymmetry while
equation \eqref{eqn:divE} can be re-expressed in terms of the gauge field
components using the relation
\begin{equation}
  E_i=F_{i0}=\partial_i A_0-\partial_0 A_i.
\end{equation}
We also impose the Lorenz gauge condition
\begin{equation}
  \label{eqn:lorenz}
  \nabla_\mu A^\mu=0
\end{equation}
to simultaneously fix the gauge and simplify the equations. 
It is expected that
a numerical solution to the equations of motion will also satisfy the constraint
equations at a given time.

\section{\label{sec:initdata} Initial Data }

To generate suitable initial data, we make a spherically-symmetric ansatz for the scalar
and gauge fields
\begin{align}
  \phi(t,\vec{x})&=f(r)e^{i\omega t},  \label{eqn:sph-ansatz-f} \\
  A_0(t,\vec{x})&=A_0(r),\\
  A_i(t,\vec{x})&=0.
  \label{eqn:sph-ansatz-A}
\end{align}
Inserting this ansatz into the equations of motion yields the following system
of coupled equations:
\begin{align}
  f''(r)+\frac{2}{r}f'(r)+f(r)g(r)^2-\frac{1}{2}\frac{d}{df}V(f)&=0,
  \label{eqn:shooting-a}\\
  A_0''(r)+\frac{2}{r}A_0'(r)+2ef(r)^2g(r)&=0.
  \label{eqn:shooting-b}
\end{align}
Here, $V(f)$ is the scalar potential and $g(r)= \omega-eA_0(r)$.  This system
constitutes an eigenvalue problem for the parameter $\omega$ subject to the
boundary conditions
\begin{align}
  \lim_{r\rightarrow \infty}f(r)=f'(0)&=0,\label{eqn:shooting-bc-a}\\
  \lim_{r\rightarrow \infty}A_0(r)=A_0'(0)&=0,\label{eqn:shooting-bc-b}
\end{align}
which are required to ensure finiteness of energy and regularity at the origin.

Gauged Q-ball solutions can be found by solving the system of equations
\eqref{eqn:shooting-a}--\eqref{eqn:shooting-b} using an iterative shooting
technique \cite{Press2007} to simultaneously determine $f(r)$ and $A_0(r)$.  In
this method, an initial choice is made for the value of $g(0)$ and a
corresponding guess is made for $f(0)$. The equations are then integrated on a
uniform grid using a fourth-order classical Runge-Kutta method out to a finite
radius $r_0$. Depending on the asymptotic behaviour of $g(r)$ and $f(r)$ at
$r_0$, the value of $f(0)$ is adjusted through iterative bisection until the
boundary conditions \eqref{eqn:shooting-bc-a} and \eqref{eqn:shooting-bc-b} are
approximately satisfied at large $r$. Once a solution is found, the eigenvalue
$\omega$ can be determined from the asymptotic value of $g(r)$ using the
boundary condition \eqref{eqn:shooting-bc-b} and $A_0(r)$ can be determined as
$A_0(r)=(\omega-g(r))/e$.

One of the main computational challenges associated with this procedure is the
high numerical accuracy required in order to find satisfactory solutions.
Typically, the number of digits required for convergence will exceed the
capacity of double-precision floating-point numbers. To overcome this
limitation, we employ the arbitrary-precision arithmetic capabilities of
\textsc{Maple} \cite{Maple}. The software precision is adjusted and the
integration is carried out until the asymptotic behaviour of $f(r)$ is observed
to decay exponentially at large $r$. At this point, the value of $f(r)$ is
typically very small (one part in $10^{8}$ or smaller) and so the fields
$g(r)$ and $f(r)$ approximately decouple in equations
\eqref{eqn:shooting-a} and \eqref{eqn:shooting-b}. In this asymptotic region, we
fit a $1/r$ tail to $g(r)$ and an exponentially-decaying tail to $f(r)$
\cite{Gulamov2015} so that the solution is determined to an arbitrarily
large radius.

\begin{figure}
  \includegraphics[width=\columnwidth]{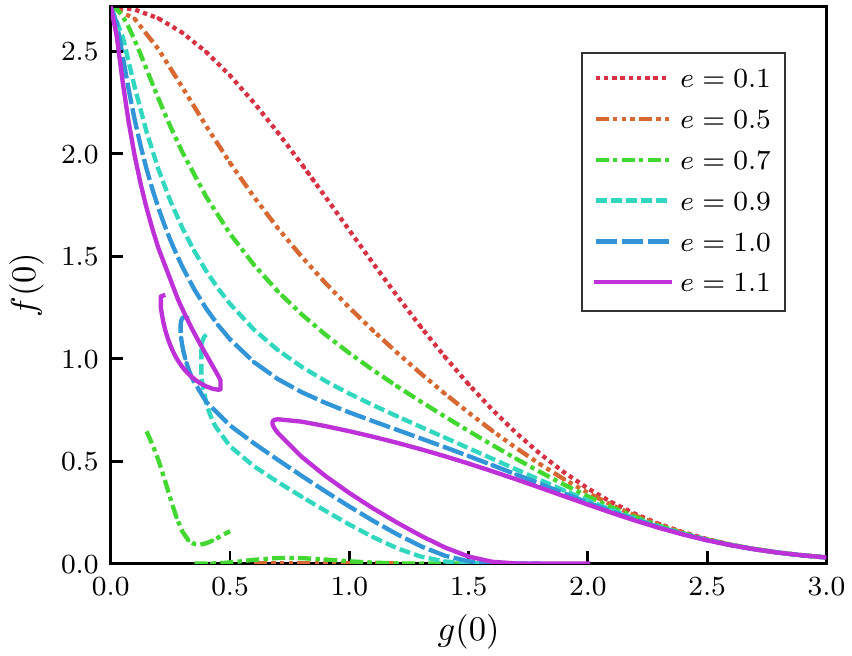}
  \caption{Shooting results for the logarithmic model \eqref{eqn:log}
  with $\mu=\beta=1$. Plotted
  is the gauged Q-ball's central scalar field value ${f}(0)$ versus
  ${g}(0)={\omega}-e{A}_0(0)$ for increasing values of
  ${e}$. Note that only unexcited gauged Q-ball solutions are presented
  here. The abrupt endpoints in the parameter space of curves with
  ${e}\ge0.7$ correspond to the appearance of additional radial nodes in
  the solution.}
  \label{fig:solnspace-log}
\end{figure}

In Fig.~\ref{fig:solnspace-log}, we present the results of our
shooting procedure for the logarithmic potential \eqref{eqn:log}. For
numerical purposes, we find it convenient to set $\mu=\beta=1$ in the
model. The central scalar field value $f(0)$ is plotted against
$g(0)={\omega}-e{A}_0(0)$ for various values of ${e}$. When the value of
$e$ is small (representing weak gauge coupling), the curve of solutions
is monotonically decreasing and single-valued, closely resembling the
behaviour of non-gauged Q-balls. However, when $e$ is
increased, the curves are no longer single-valued and they begin to
bifurcate with some curves ending abruptly in the solution space. These
distinct endpoints generally correspond to the appearance of additional
radial nodes in the solution, representing excited gauged Q-balls
\cite{Friedberg1976,Volkov2002,Kleihaus2005,Mai2012,Almumin2022}. Also
notable is the appearance of distinct curves close to the horizontal
axis where $f(0)$ is very small. These solutions correspond to Q-shells
\cite{Arodz2009,Tamaki2014,Heeck2021b} which attain their maximal field
values away from $r=0$ and resemble shell-like concentrations of the
fields.

\begin{figure}
  \includegraphics[width=\columnwidth]{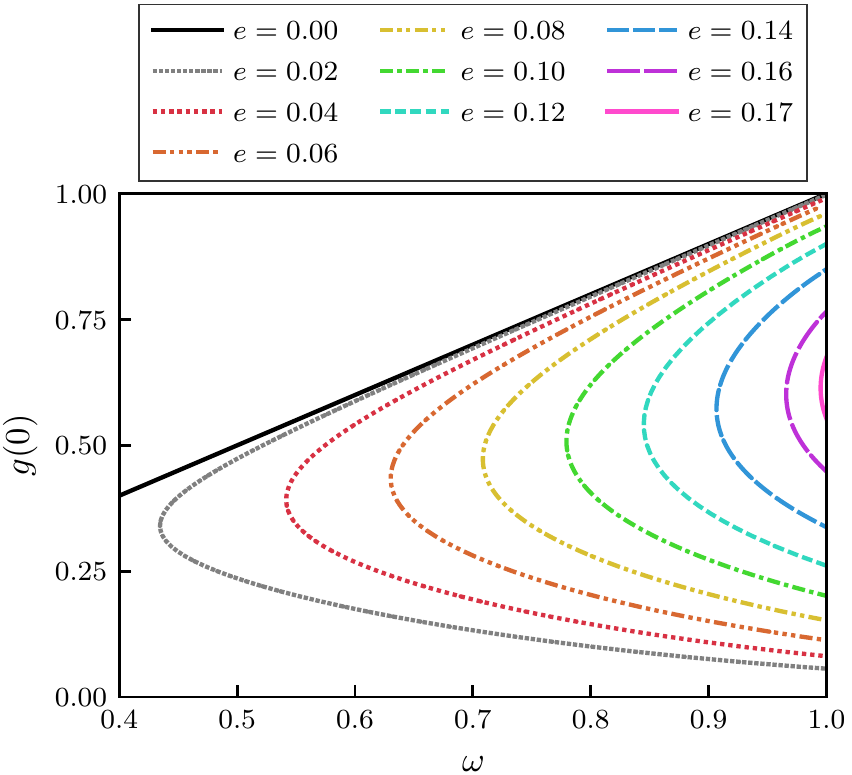}
  \caption{
    Shooting results for the polynomial model \eqref{eqn:poly} with
    $m=k=1$ and $h=0.2$. Plotted is
    ${g}(0)={\omega}-{e}{A}_0(0)$ versus the eigenfrequency
    ${\omega}$ for various values of ${e}$. A linear dependence can be 
    observed for ${e}=0.0$ (representing non-gauged Q-balls). For
    ${e}>0$, the curve bifurcates into an upper and lower branch. As the
    value of ${e}$ is increased, the range of the solutions decreases
    significantly.}
  \label{fig:solnspace-poly}
\end{figure}

Plotted in Fig.~\ref{fig:solnspace-poly} are the results of our
numerical shooting procedure for the polynomial potential
\eqref{eqn:poly} with $m=k=1$ and $h=0.2$. In order to clearly
distinguish the curves, and following Ref.~\cite{Loginov2020}, we plot
${g}(0)={\omega}-{e}{A}_0(0)$ versus ${\omega}$ for various values of
${e}$.  We restrict our shooting to solutions where ${\omega}\le
1$, which is required in order to ensure that the solutions have finite
energy \cite{Lee1989,Gulamov2015,Loginov2020}. The case of ${e}=0.0$
(corresponding to non-gauged Q-balls) is represented by a linear line in
the solution space. As ${e}$ is increased, a minimal value
${\omega}_\text{min}$ appears which separates each curve into an upper
and lower branch.  The value of ${\omega}_\text{min}$ increases with
${e}$ until ${\omega}_\text{min}>1$, at which point no gauged
Q-ball solutions can be found in the model. 
We note that while Q-shell solutions are known to exist for the polynomial
potential \cite{Heeck2021b}, no such solutions are found for our choice of the
potential parameters.

As a basic check of our shooting procedure, we have compared our numerical
solutions to those reported in previous publications on $U(1)$ gauged Q-balls
in logarithmic and polynomial models \cite{Panin2017,Loginov2020}.
We find good agreement with the previously reported results.

In order to generate initial data that is suitable for evolution in
axisymmetry, we transform the spherical solutions in \eqref{eqn:sph-ansatz-f}--\eqref{eqn:sph-ansatz-A} to
cylindrical coordinates by performing a
fourth-order polynomial interpolation of the spherical solution
in the $\rho$--$z$ plane. This provides axisymmetric initial data that will
subsequently be used in our numerical simulations.

\section{\label{sec:diagnos} Diagnostics}

Here we discuss several diagnostics which are useful in characterizing the
stability of each gauged Q-ball configuration. For the purposes of this work, a
configuration is defined to be stable if small perturbations to the initial
state remain bounded during the course of the evolution. Unstable
configurations are those for which small perturbations grow exponentially on
top of the solution, eventually leading to the destruction of the Q-ball 
(such as through fragmentation or dispersal of the fields). Note that with this
definition, we classify as stable those configurations for which the fields may
be weakly oscillating or radiating but are not destroyed by the initial perturbation.

There are several physical quantities associated with the scalar and gauge
fields which are relevant when monitoring the evolution of each configuration.
Principal among these are the conserved Noether charge $Q$ and the total energy
$E$. The Noether charge is given by
\begin{equation}
  \begin{split}
    Q=2\pi i \iint d\rho\, dz\, \rho\, \bigl [ \phi(D^0\phi)^*-\phi^*
    (D^0\phi) \bigr ]
    \end{split}
\end{equation}
and the total energy is given by
\begin{equation}
  \begin{split}
    E = 2\pi\iint d\rho&\, dz\, \rho\, \biggl [ F_{0\alpha}F_0^{\;\alpha} + \frac{1}{4}F_{\alpha\beta}F^{\alpha\beta} + V(|\phi|)\\
    &+(D_\alpha\phi)(D^\alpha\phi)^*+ 2 (D_0\phi) (D_0\phi)^* \biggr ] .
    \end{split}
\end{equation}
Both $Q$ and $E$ are time-independent quantities which are expected to be
conserved as long as the fields remain localized within the simulation domain.

We will now discuss how we add small perturbations to the
stationary initial data. The solutions are perturbed in two ways: (i)
perturbation through the inherent numerical truncation error of the
finite-difference scheme, and (ii) perturbation by an auxiliary scalar field
designed to explicitly excite all underlying modes of the configuration.

\subsection{Perturbation by Numerical Truncation Error}

As a first test of the stability of our gauged Q-ball configurations, we
numerically evolve the fields forward in time using the axisymmetric initial
data described in Sec.~\ref{sec:initdata}.  Upon evolution, the fields will
be subject to small numerical perturbations due to the finite-difference
discretization which is used to solve the evolution equations.  This can be
understood from the observation that the discrete solution of a uniform
centered finite-difference scheme admits a truncation error expansion around
the continuum solution in powers of the grid spacing \cite{Richardson1911}.
While the exact form of this expansion is generally not known (making the
perturbations effectively random), the magnitude of the associated truncation
error is tied closely to the numerical resolution of the simulation and can
therefore be indirectly controlled. In the sections that follow, we will refer
to perturbations of this form as ``Type 0''.

One consequence of this type of perturbation is that any potential
instabilities will take longer to manifest for higher-resolution
simulations than for lower resolution ones.  This is because the
magnitude of the truncation error becomes smaller at higher resolutions.
It is therefore necessary to evolve the configuration over sufficiently
long times in order to assess stability.  
The notion of a ``sufficiently long time" is difficult to make precise,
but this timescale can generally be estimated by observing the
oscillations of the scalar field modulus $|\phi|$ when subject to a
perturbation. Even for small perturbations, the maximum value of
$|\phi|$ will tend to oscillate at frequencies which correspond to the
underlying modes of the configuration. The dynamical time can then be
identified as the inverse frequency of the longest mode. For
the problem at hand, we evolve each configuration with this timescale in
mind so that any slowly-growing unstable modes have the opportunity to
manifest.  We find that this procedure provides an adequate preliminary
test of stability which can be further verified using additional
perturbation methods (to be discussed immediately below).

\subsection{Perturbation by an Auxiliary Scalar Field}

As a second test of stability, we dynamically perturb the gauged Q-balls by
simulating the implosion of an asymmetric shell of matter onto the stationary
configurations.  We do this by introducing a massless real scalar field
$\chi(t,\rho,z)$ that couples to the complex Q-ball 
field $\phi(t,\rho,z)$ in
the modified theory:
\begin{equation} \label{eqn:gauged-lagr-pert}
\begin{split}
  \mathcal{L} = -\left(D_\mu\phi\right)^* D^\mu\phi &-
  V\left(|\phi|\right)-\frac{1}{4}F_{\mu\nu}F^{\mu\nu}\\&-\partial_\mu\chi\partial^\mu\chi
  - U(|\phi|,\chi),
\end{split}
\end{equation}
where $U(|\phi|,\chi)$ describes the interaction potential of the fields $\phi$
and $\chi$.  We compute the modified equations of motion from
\eqref{eqn:gauged-lagr-pert} to obtain
\begin{align}
  -\nabla_\mu F^{\mu\nu}-ie\phi(D^\nu\phi)^*+(ie\phi^*)D^\nu\phi&=0, \label{eqn:eom-pert-a}\\
  D_\mu D^\mu \phi - \frac{\partial}{\partial
  \phi^*}V(|\phi|)-\frac{\partial}{\partial \phi^*}U(|\phi|,\chi)&=0,\label{eqn:eom-pert-b}\\
    \nabla_\mu\nabla^\mu\chi-\frac{1}{2}\frac{\partial}{\partial\chi}U(|\phi|,\chi)&=0. \label{eqn:eom-pert-c}
\end{align}
One can see from \eqref{eqn:eom-pert-a}--\eqref{eqn:eom-pert-c} that by
choosing an interaction potential $U(|\phi|,\chi)$ such that
$U(|\phi|,\chi)\rightarrow 0$ in the limit of $\chi\rightarrow 0$, then the
modified equations \eqref{eqn:eom-pert-a} and \eqref{eqn:eom-pert-b} reduce to
equations \eqref{eqn:eom-a} and \eqref{eqn:eom-b} (with \eqref{eqn:eom-pert-c}
just representing an independent wave equation for $\chi$). This means that
$\chi$ and $\phi$ will elicit some mutual influence when the fields overlap,
but the influence will disappear if the fields become well-separated. In this
sense, $\chi$ can act as an external perturbing agent. We initialize $\chi$ as
an ingoing asymmetric shell of the form
\begin{equation}
   \chi(0,\rho,z)=A\exp\left[-\left(\frac{\sqrt{\frac{(\rho-\rho_0)^2}{a^2}+\frac{(z-z_0)^2}{b^2}}-r_0}{\delta}\right)^2\right],
\end{equation}
\begin{equation}
   \partial_t\chi(0,\rho,z)=\frac{\chi+\rho\partial_\rho\chi+z\partial_z\chi}{\sqrt{\rho^2+z^2}},
\end{equation}
where $A$, $a$, $b$, $\delta$, $r_0$, $\rho_0$ and $z_0$ are parameters
specifying the characteristics of the initial pulse. If $r_0$ is made large, the
field approximately vanishes in the vicinity of the gauged Q-ball at the initial
time and so $\chi$ can be considered an external perturbation with
a size controlled by $A$.
Strictly speaking, the notion of an ``external"
perturbation cannot be made precise because gauged Q-balls do not have a finite
radius. However, since the scalar field decays exponentially away from the
center of the configuration, initializing the auxiliary field sufficiently far
away from the center will serve as a good approximation to an external
perturbation. Note also that the auxiliary field couples only to the scalar
field so that the long-range behaviour of the gauge field is not a significant factor.

During the evolution, the massless scalar field implodes toward the origin and
collides with the gauged Q-ball. The two fields temporarily interact before the
massless field passes through the origin and explodes outward to $r\rightarrow
\infty$, leaving the gauged Q-ball perturbed at the origin.  Due to the
asymmetry of the imploding pulse, the interaction of the two scalars is
expected to excite the underlying modes of the system and induce the 
disruption of
any unstable configurations.  For our purposes, we choose
\begin{equation}
U(|\phi|,\chi)=c|\phi|^2\chi^2
\end{equation}
where $c$ is a coupling constant that
controls the coupling strength between $\chi$ and $\phi$. In the sections that
follow, we will refer to perturbations of this form as ``Type I''. 
This technique
resembles the methods of Ref.~\cite{Hawley2000} to investigate the stability of
boson stars.

We note that configurations which are subject to perturbations of this type
will inevitably also be perturbed by the inherent truncation error of the
numerical simulation (Type 0). However, since Type 0 perturbations are
typically very small and effectively random, Type I perturbations provide an
additional level of control in determining the stability of a given
configuration. For the results presented here, the simulations are repeated for
various values of the Type I perturbation parameters $A$ and $c$. This is done
to verify that the response of the Q-ball field to the perturbation (as
measured, for example, by the magnitude of the induced oscillations of the
scalar field modulus $|\phi|$) remains in the linear regime: an increase of $A$
or $c$ leads to a corresponding increase in the magnitude of
oscillations of the perturbed $|\phi|$.

\section{\label{sec:results} Numerical Results}

Here we present results from the dynamical evolution of gauged Q-balls in the
potentials \eqref{eqn:log} and \eqref{eqn:poly}. For each simulation, we
numerically solve the evolution equations using a second-order Crank-Nicolson
finite-difference scheme implemented using the Rapid Numerical
Prototyping Language (\textsc{RNPL}) framework
\cite{Marsa1995}.  Fourth-order Kreiss-Oliger dissipation is applied as a mild
low-pass filter to 
damp poorly-resolved and potentially problematic (from a numerical 
stability viewpoint) high-frequency solution components.
We implement a modified
Berger-Oliger adaptive mesh refinement (AMR) algorithm via the
\textsc{PAMR/AMRD} libraries \cite{Pretorius2006} to increase the numerical
resolution of our simulations. In all examples presented below, the base grid is
taken to be $129\times257$ grid points in $\{\rho,z\}$ and up to five additional levels
of mesh refinement are used with a refinement ratio of 2:1. The domain is taken
to be finite with outgoing boundary conditions imposed at the outer boundaries.
Reflective (or anti-reflective) boundary conditions are imposed at the axis of
symmetry in order to ensure regularity of the fields. 
We choose a Courant factor of $\lambda=dt/\min\{d\rho,dz\}=0.25$ and
evolve the configurations until at least $t\approx 1000$
to assess stability, though in many cases we evolve for longer in order to observe the late-time dynamics. Further details about the numerical
implementation and code validation are given in
App.~\ref{apx:numerics}.

To illustrate the general behaviour of stable and unstable configurations, we
focus on several specific solutions for the potentials \eqref{eqn:log} and
\eqref{eqn:poly}. These solutions are listed in Table \ref{table:evos}. 
Configurations L1--L4 correspond to the logarithmic potential while 
configurations P1--P2 correspond to the polynomial potential.  We note that
besides L1--L4 and P1--P2, we have also performed hundreds of additional
evolutions along the solution curves of Fig.~\ref{fig:solnspace-log} and
Fig.~\ref{fig:solnspace-poly} in order to determine the general regions of
stability. This stability investigation  will be discussed below.

{ \setlength\extrarowheight{2pt} \begin{table*}
  \begin{tabular}{|c|c|c|c|c|c|c|c|c|c|c|c|} \hline
    Configuration & Result   & $\phi(0,0)$           & $ A_0(0,0)$  & $e$   & $\omega$  & $E$                  & $Q$                   & $d_{max}$    & Perturbation  & $c$  & $A$  \\ \hline \hline
    L1            & stable   & 0.6461                & 1.383        & 1.1   & 2.522     & 52.08                & $-22.37$              & 50           & Type I        & 0.1  & 0.1  \\
    L2            & unstable & 1.179                 & 3.159        & 1.1   & 3.695     & 281.5                & $-94.34$              & 50           & Type 0        & --   & --   \\
    L3            & unstable & $2.448\times10^{-13}$ & 0.9803       & 1.1   & 3.078     & 260.3                & $-92.76$              & 75           & Type 0        & --   & --   \\
    L4            & unstable & 1.539                 & 2.254        & 1.1   & 2.680     & 95.18                & $-38.13$              & 50           & Type 0        & --   & --   \\
    P1            & stable   & 1.973                 & 2.515        & 0.17  & 0.9976    & 405.1                & $-387.5$              & 50           & Type I        & 0.1  & 0.1  \\
    P2            & unstable & 1.904                 & 46.94        & 0.02  & 0.9958    & $1.076\times10^{6}$  & $-1.480\times10^{6}$  & 150          & Type 0        & --   & --   \\ \hline
  \end{tabular} \caption{ Results for several representative gauged Q-ball
  evolutions. The configurations L1--L4 correspond to solutions found using the
  logarithmic potential \eqref{eqn:log}. P1--P2 represent configurations found
  using the polynomial potential \eqref{eqn:poly}. The second column indicates
  the outcome of the numerical evolution. From left to right, the remaining
  columns give the initial central scalar field amplitude $\phi(0,0)$, the
  initial central gauge field value $A_0(0,0)$, the electromagnetic coupling constant
  $e$, the eigenfrequency $\omega$,
  the total integrated energy $E$, the total Noether charge $Q$, the size of
  the simulation domain spanning $\{\rho\,:\,0\le\rho\le d_{max}\}$ and
  $\{z\,:\,-d_{max}\le z \le d_{max}\}$, the type of perturbation used, and the
perturbation parameters $c$ and $A$ (if applicable).}
\label{table:evos} \end{table*} }

\subsection{$V_\text{log}$ Model}

Here we consider the dynamical stability of gauged Q-balls in the
logarithmic model \eqref{eqn:log}. Most relevant for this work are the
results of Ref.~\cite{Panin2017} which conducted numerical evolutions of
gauged Q-ball configurations for ${e}=1.1$ in spherical symmetry. There
it was found that both stable and unstable gauged Q-balls can exist in
the model, though the classical stability criterion
\eqref{eqn:VK-criteria} provides little information about the stability
of a given configuration in the general case.  Once again, we set
$\mu=\beta=1$ for numerical purposes. For brevity, and to facilitate
comparison with previous work, we focus on the case of ${e}=1.1$
where the system is fully coupled.

\begin{figure}
\includegraphics[width=\columnwidth]{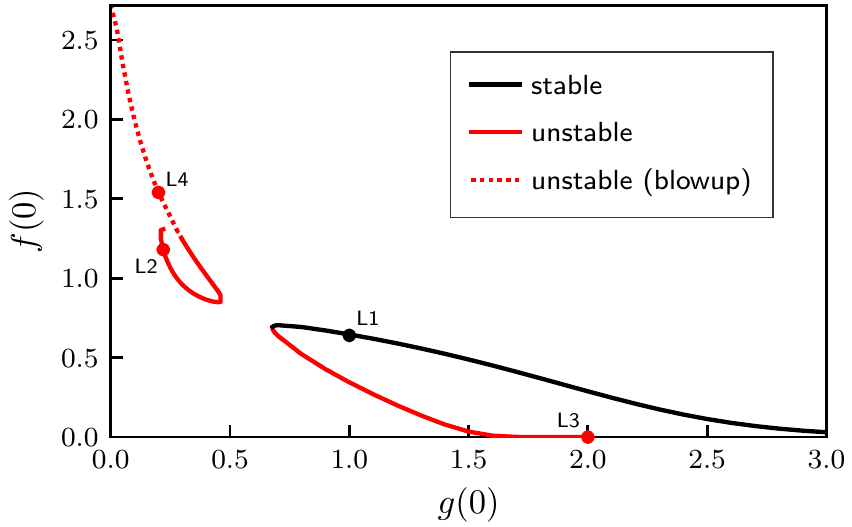}
  \caption{\label{fig:solnspace-log-stab} Regions of stability and instability
  for gauged Q-balls in the $V_\text{log}$ model with ${e}=1.1$. The
  points L1--L4 correspond with the configurations listed in Table
  \ref{table:evos}. The solid black line represents stable configurations while
  red lines represent regions of instability. The dashed red line indicates
  regions where blowup of the solutions is observed; see the main text for
  details.
  }
\end{figure}

\begin{figure}
\includegraphics[width=\columnwidth]{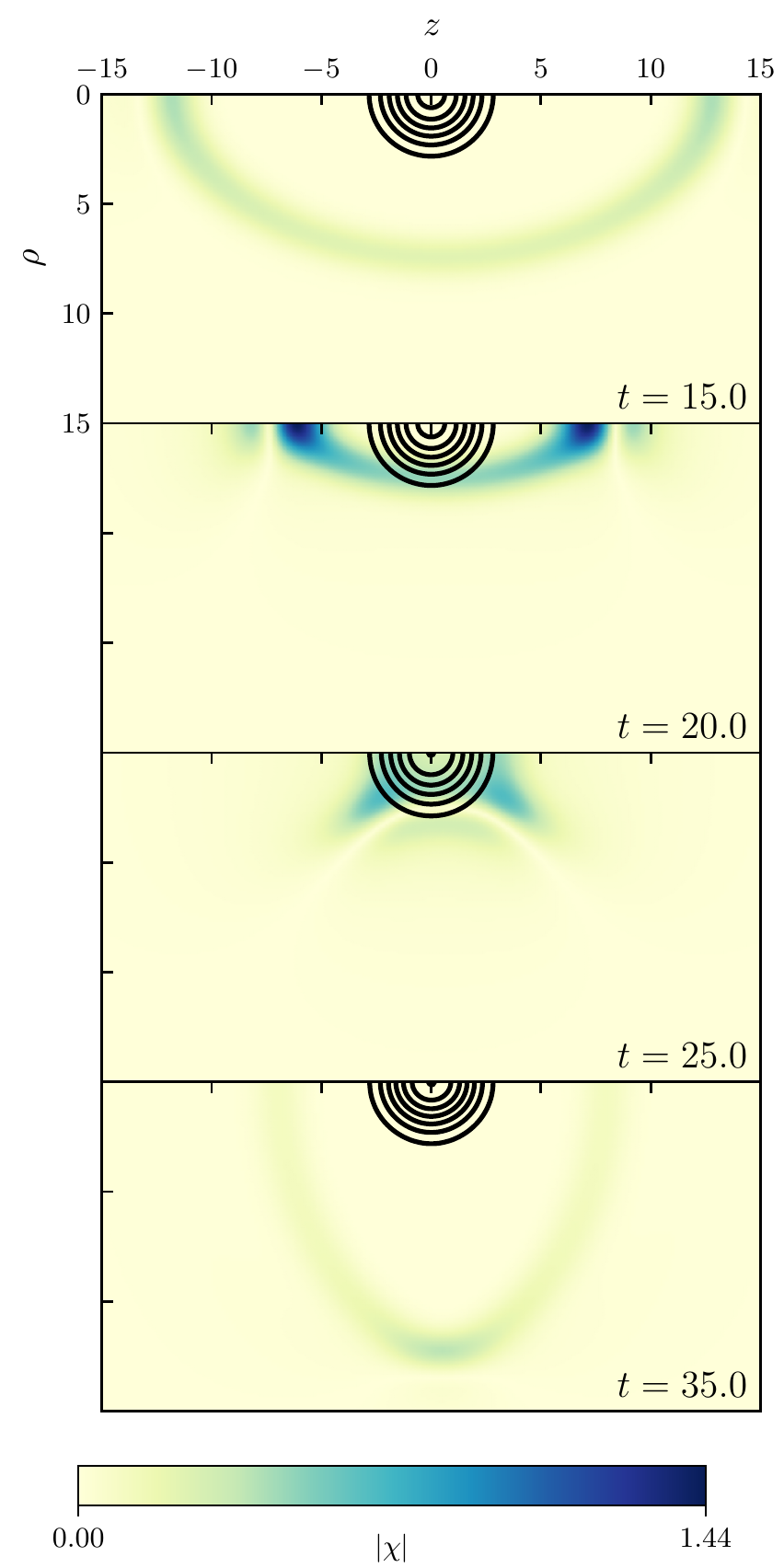}
  \caption{\label{fig:L1-evo} Magnitude $|\chi|$ of the perturbative scalar
  field interacting with a stationary gauged Q-ball (represented by contours)
  corresponding to configuration L1 in Table \ref{table:evos}. The contour
  lines represent the Q-ball field modulus $|\phi|$ in steps of 0.1. The axis
  of symmetry is coincident with the top edge of each panel. The simulation
  domain spans $\{\rho\,:\,0\le\rho\le 50\}$ and $\{z\,:\,-50\le z \le 50\}$.
  }
\end{figure}

The relevant properties of each of the configurations L1--L4 are described in
Table \ref{table:evos} along with the final result of numerically evolving the
configuration forwards in time. In Fig.~\ref{fig:solnspace-log-stab}, the
location of each of these configurations in the solution space is labelled with
a dot. L1 corresponds to a solution on the stable branch. L2 corresponds to an
unstable configuration which decays through dissipation of the fields. L3
corresponds to an unstable Q-shell solution which breaks apart into several
smaller solitonic components. Finally, L4 illustrates the case of an unstable
solution which responds to perturbation by growing without bound. Here, only L1
is subject to the Type I perturbation (to illustrate the dynamical stability of
the configuration) while L2--L4 are all subject to Type 0 perturbations only.

\begin{figure}
\includegraphics[width=\columnwidth]{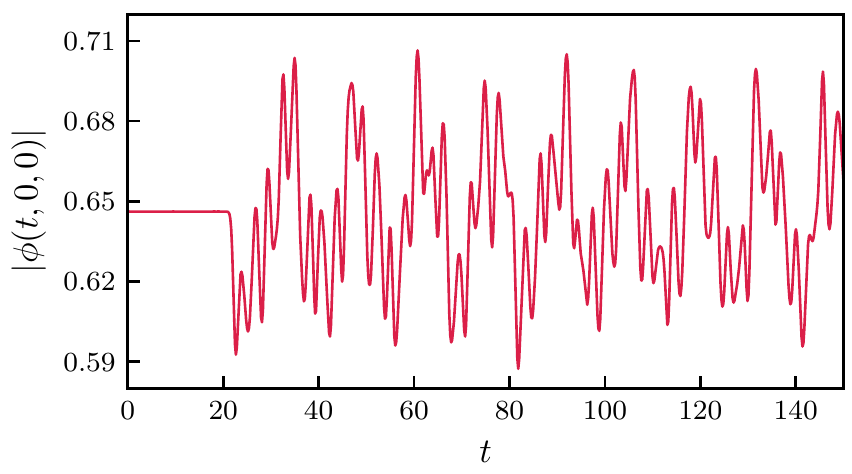}
  \caption{\label{fig:L1-osc} Oscillations of the scalar field modulus $|\phi|$
  for configuration L1 subject to a Type I perturbation with parameters
  $c=0.1$ and $A=0.1$. Corresponding snapshots for the evolution are given in
  Fig.~\ref{fig:L1-evo}. The perturbative field $\chi$ implodes upon the
  Q-ball at $t\approx20$, causing large oscillations in $|\phi|$. Over the full timescale
  of the evolution, the Q-ball slowly returns toward the original configuration.} 
\end{figure}

First let us discuss the evolution of configuration L1. This evolution is run
for 65000 base-grid time steps up to a final time of $t=6400$.
To assess the
stability, we apply a Type I perturbation with parameters $c=0.1$ and
$A=0.1$. Results for this evolution are shown in Fig.~\ref{fig:L1-evo}. The
contour lines in the figure represent the scalar field modulus $|\phi|$ while
the colormap represents the perturbing field modulus $|\chi|$. The imploding
pulse, which is centered around the point $\{\rho=0.0,z=0.5\}$, interacts
with the Q-ball starting at $t\approx20$ (the second panel of the figure)
and explodes through the origin, leaving the simulation domain at
$t\approx70$.  This induces small asymmetric distortions in the Q-ball
field which can be observed as changes of the contour lines in the
subsequent panels. This distortion also creates large oscillations in the
maximal value of $|\phi|$ which are plotted in Fig.~\ref{fig:L1-osc}.  Prior
to the imploding pulse interacting with the Q-ball, the oscillations of
$|\phi|$ are very small and are sourced by Type 0 perturbations. After the
pulse interacts with the Q-ball at $t\approx20$, the amplitude of the
oscillations grows significantly as the imploding pulse transfers energy to the
configuration.  It oscillates continuously around the stationary (unperturbed)
solution before slowly returning toward the original configuration.

If configuration L1 was unstable, one would expect that the interaction
between $\phi$ and $\chi$ would excite any unstable modes underlying the
solution. Once excited, these modes should quickly grow and bring about the
destruction of the configuration. However, no such behaviour is observed in our
numerical experiments using different values of $c$ and $A$. In addition, we
have also analyzed the behaviour of the configuration when subject to Type 0
perturbations only, finding no evidence of instability.
We therefore conclude that L1 is stable.

\begin{figure} [t]
\includegraphics[width=\columnwidth]{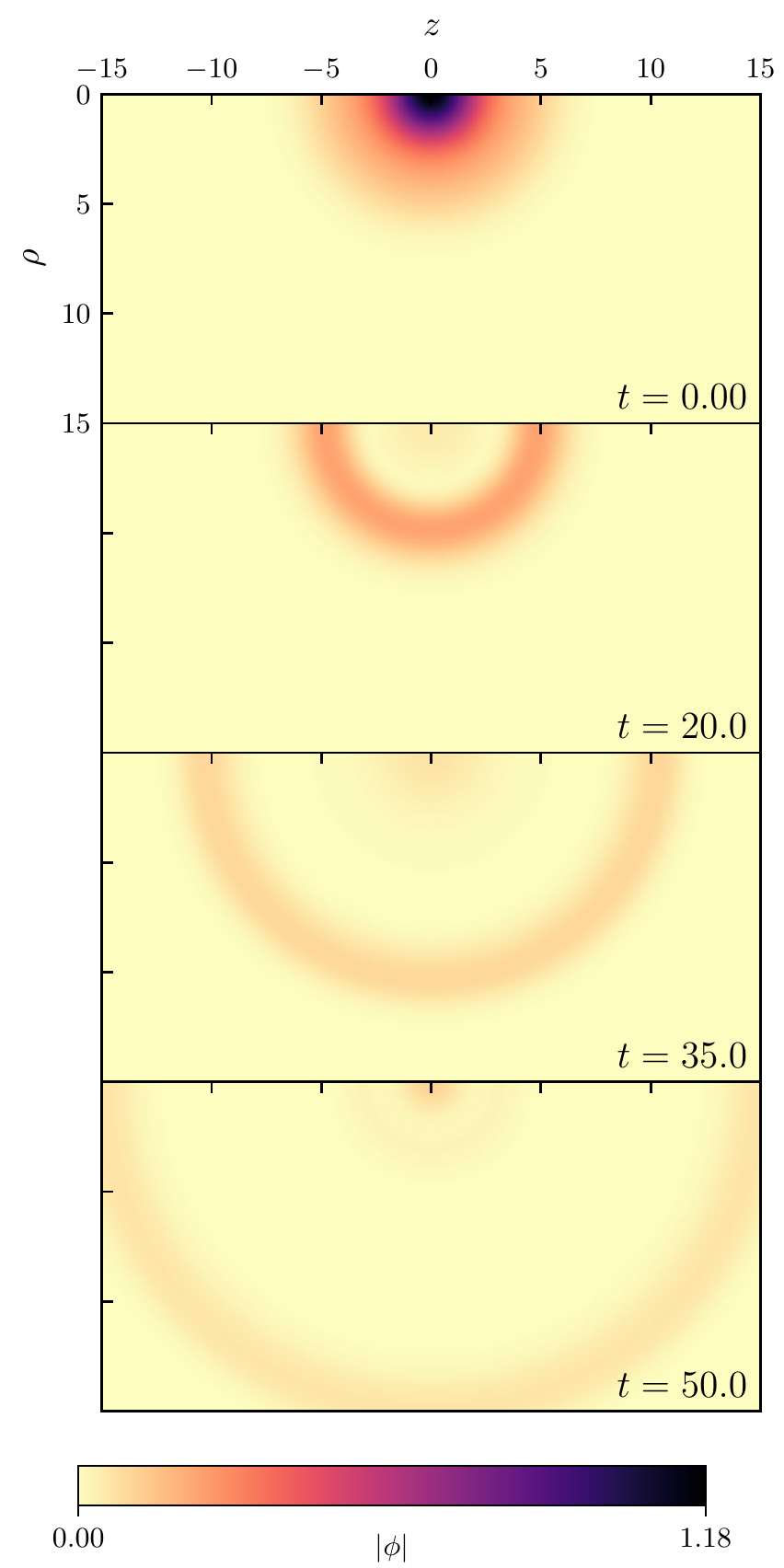}
  \caption{\label{fig:L2-evo} Evolution of the scalar field modulus $|\phi|$
  for configuration L2 subject to a Type 0 perturbation. Upon evolution,
  the gauged Q-ball rapidly decays until no significant remnant of the initial
  configuration remains in the simulation domain. The axis of symmetry is
  coincident with the top edge of each panel. The simulation
  domain spans $\{\rho\,:\,0\le\rho\le 50\}$ and $\{z\,:\,-50\le z \le 50\}$.}
\end{figure}

Next we consider L2, which lies on the upper branch of the solution curve in
Fig.~\ref{fig:solnspace-log-stab}. We subject this configuration only to a
Type 0 perturbation. The time evolution of L2 is depicted in 
Fig.~\ref{fig:L2-evo}. 
The scalar field modulus retains its initial shape for only a
short time before quickly decaying and spreading radially. As the evolution
proceeds, the fields continue to propagate toward the boundaries until no
significant remnant of the initial configuration remains in the domain. As
mentioned previously, the timescale over which the Q-ball survives before
dissipating can be extended by increasing the numerical resolution of the
simulation (thereby decreasing the size of the perturbation).  However, even
with five additional levels of refinement, the solution begins to decay within the first
few oscillations of the scalar field. Since the outcome of this evolution is
the total dispersal of the fields, we classify L2 as unstable.

\begin{figure}[t]
\includegraphics[width=\columnwidth]{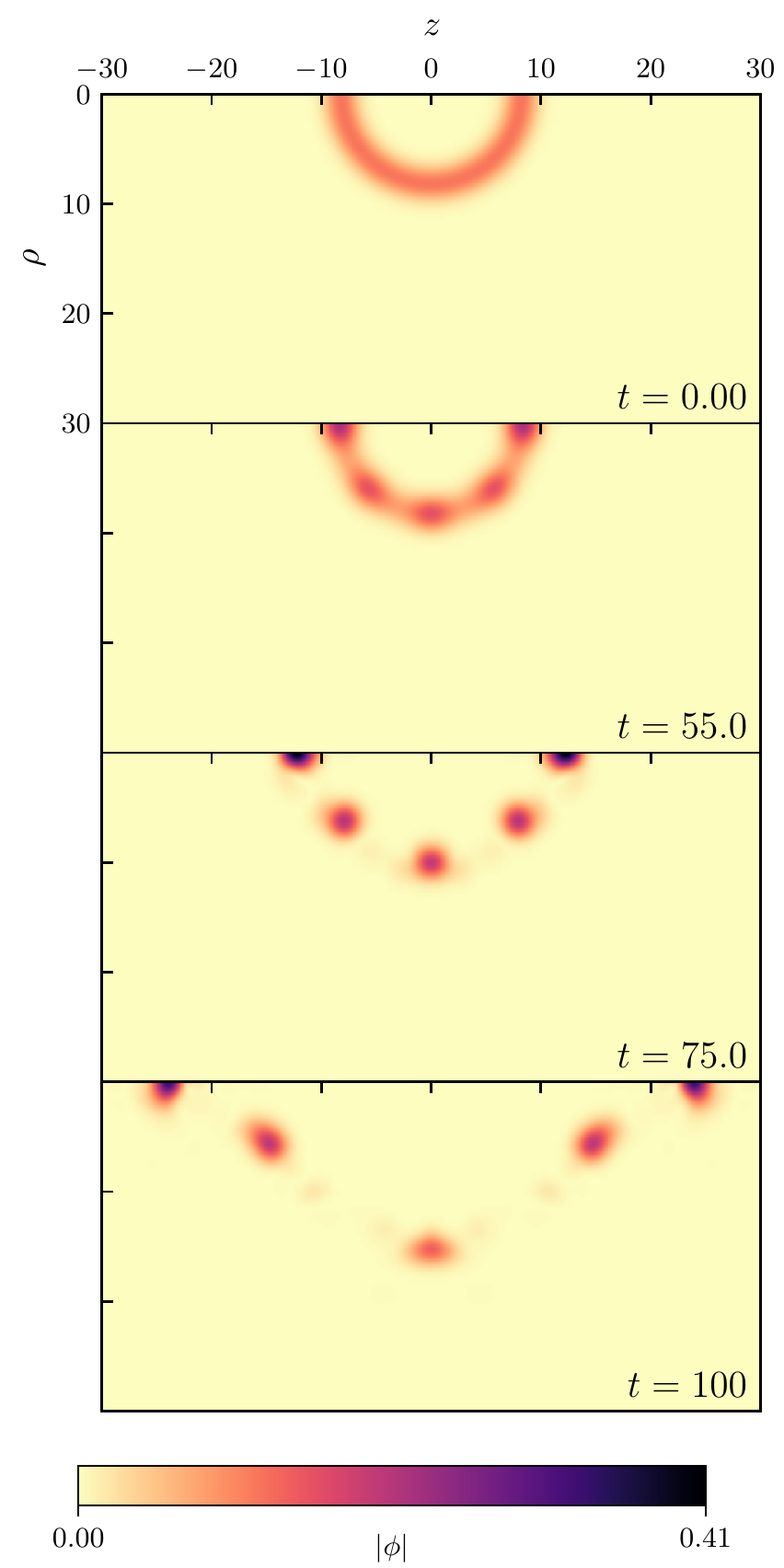}
  \caption{\label{fig:L3-evo} Evolution of the scalar field modulus $|\phi|$
  for configuration L3 subject to a Type 0 perturbation. Notable in this evolution is
  the formation of non-spherical ring-like structures which coalesce away from
  the axis of symmetry and can survive for some time. The axis
  of symmetry is coincident with the top edge of each panel. The simulation
  domain spans $\{\rho\,:\,0\le\rho\le 75\}$ and $\{z\,:\,-75\le z \le 75\}$.
  }
\end{figure}

Consider next L3, which lies on the lower branch of
Fig.~\ref{fig:solnspace-log-stab}, near the bottom axis. Notably,
${f}(0)\approx 0$ near this axis which results in $|\phi|$ attaining its
maximal value away from the origin. This configuration resembles a shell-like
distribution of matter and is therefore labelled a ``gauged Q-shell". The
evolution of this configuration subject to a Type 0 perturbation is shown in
Fig.~\ref{fig:L3-evo}. At the initial time (top panel), the shell-like nature
of the solution is readily apparent. As time evolves, the spherical symmetry of
the configuration is quickly broken as the Q-shell fragments into several
individual components which propagate away in different directions. Two of
these components remain centered on the axis of symmetry and travel along this
axis toward the outer boundaries. These components remain approximately
spherical for the entirety of the evolution (aside from oscillations and
distortions induced by the fragmentation process) and represent smaller
``child" gauged Q-balls of the initial configuration. In addition, we observe
that the field also fragments into several distinct components which coalesce
away from the axis of symmetry. In three-dimensions, these resemble ring-like
structures which we call ``gauged Q-rings". Those Q-rings which are closest to
the axis quickly collapse back into spherical structures (child gauged Q-balls)
which remain on the axis of symmetry for the rest of the evolution. However,
those rings which are initially farthest away from the axis of symmetry are
observed to propagate outward and can survive for some time.  The bottom panel
of Fig.~\ref{fig:L3-evo} illustrates the behaviour of the gauged Q-rings
associated with the decay of L3. The largest Q-ring  reaches a maximal distance
from the origin of $\rho\approx40$ before turning around and collapsing back
onto the axis of symmetry by $t\approx500$.

We classify L3 as an unstable configuration. Moreover, we find that all Q-shell
solutions on the lower branch of Fig.~\ref{fig:solnspace-log-stab} are
unstable. It is notable that this particular branch of solutions was
reported to be stable in Ref.~\cite{Panin2017} under spherical symmetry
assumptions.  
However, the formation of rings is obviously forbidden under
spherical symmetry, so our current results are not inconsistent with 
previous findings.

The formation of gauged Q-rings does not appear to be a unique feature of the
decay of L3. We observe a similar phenomenon for other Q-shells on the lower
branch of Fig.~\ref{fig:solnspace-log-stab} as well as for the decay of some
unstable gauged Q-balls on the upper branch (though the resulting rings may
differ in size and lifetime).  We have not been able to find any gauged Q-rings
which can survive indefinitely. In each case, the rings propagate outward some
finite distance from the axis of symmetry before collapsing back inward and
forming a gauged Q-ball.  This behaviour is similar to what has been observed
for non-gauged Q-balls. In Ref.~\cite{Battye2000}, rings are formed through the
high-energy collisions of non-gauged Q-balls which also collapse back inward
at late times. Q-ring solitons with semitopological origin have also been
considered in Ref.~\cite{Axenides2001}. While the rings observed here do not persist
indefinitely, they appear to retain their shape despite the relatively violent
dynamics that occur after the decay of L3 (until eventually collapsing onto the
axis of symmetry).
Since these rings could potentially survive long enough to
interact with other structures and produce dynamical effects, we conjecture
that they may represent a new type of non-spherical solution
in the model. 

\begin{figure}
    \includegraphics[width=\columnwidth]{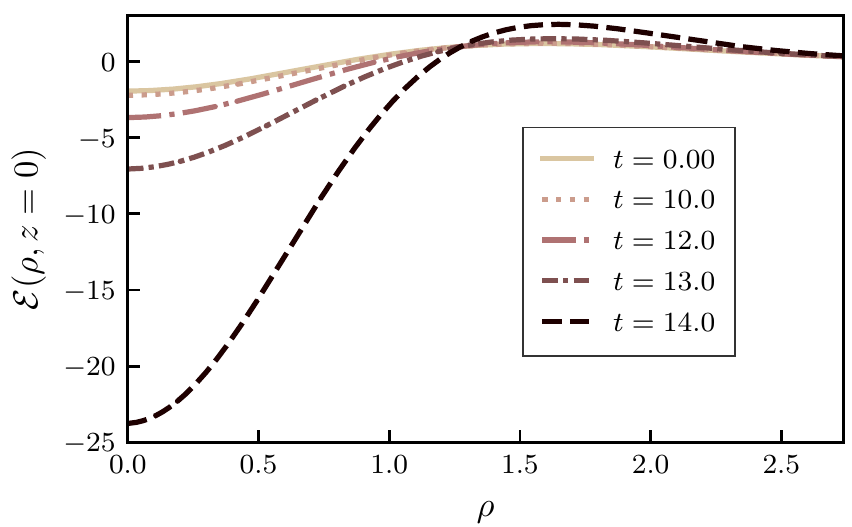}
    \caption{\label{fig:L4-Eden} Radial slices of the energy density
    $\mathcal{E}$ of L4 along $z=0$ at various times during the
    evolution. Initially, the energy density is negative in a region
    surrounding the origin and positive elsewhere. As the evolution proceeds,
    the energy density near the origin grows; the positive region also grows to
    compensate. Note that the total energy integrates to a positive quantity
    and is conserved to within 1\% over the timescales shown here.}
\end{figure}

Finally, let us discuss the evolution of L4. This configuration lies on the
upper branch of Fig.~\ref{fig:solnspace-log-stab} above L2. This
configuration is subject only to the Type 0 perturbation. When evolved forward
in time, we observe that the modulus of the scalar field quickly grows without
bound until large field gradients are produced. These excessive field gradients
cannot be numerically resolved by our code even with increasing adaptive mesh
refinement, leading to termination of the evolution due to computational
constraints. In Fig.~\ref{fig:L4-Eden}, we plot a radial slice of the energy
density of the configuration at various points during the evolution. At the
initial time, the energy density of the configuration is already negative near
the origin.  This is likely a consequence of the logarithmic scalar field
potential \eqref{eqn:log} being unbounded from below: when the value of
$|\phi|$ is large enough, the scalar potential term $V(|\phi|)$ in
\eqref{eqn:emt} can become negative. If $V(|\phi|)$ dominates locally over the
other energies in the system, then the energy density at a point in space can
also become negative (even while the total integrated energy remains overall
positive).  When this configuration is evolved forward in time, it may become
energetically favourable for the field in the negative region to grow.
At the same time, the field in regions of positive energy density would have to
grow to compensate and keep the total integrated energy constant.  This runaway
process results in the large field gradients and unbounded growth (blowup)
observed in Fig.~\ref{fig:L4-Eden}.

While the decay process of L4 may be unphysical, it is not entirely unexpected.
Similar phenomena have been observed in other Q-ball models where local energy
densities can attain negative values \cite{Anderson1970,Panin2019}.  It is also
possible that the decay of L4 could manifest in a different manner (such as
dissipation of the fields, similar to L2) if the sign of the initial
perturbation to the system could be precisely controlled. However, this level
of control is not possible with the Type 0 perturbation, and the fields are
found to grow too quickly for Type I perturbations to be effective. In any
case, the evolution of L4 results in the destruction of the configuration, and
we therefore classify L4 as unstable.

Having discussed the specific configurations L1--L4, we now turn to the general
regions of stability and instability depicted in 
Fig.~\ref{fig:solnspace-log-stab}. The black solid line on the bottom branch
indicates the regions of the solution curve which are found to be stable under
both Type 0 and Type I perturbations. L1 lies in this region.
At the leftmost edge of the bottom branch, we observe a
turning point where the gauged Q-ball configurations suddenly become unstable.
As this turning point is approached from above, the stable gauged Q-ball
configurations become less robust: it becomes possible for a sufficiently large
Type I perturbation to ``kick" the configuration to the unstable branch,
though the same configuration remains stable for smaller-sized perturbations.
Due to this effect, it is difficult to exactly determine the location of the
onset of instability. However, our numerical experiments suggest that it
corresponds with the leftmost edge of the lower branch as depicted in the
figure.  The region of the curve below this point, marked by a red solid line,
is found to be unstable. This region contains L3 along with other Q-shell
solutions.  Lastly, all portions of the curve along the upper branch
(containing L2 and L4) are found to be unstable.  The solutions which are found
to exhibit the blowup behaviour when subject to Type 0 perturbations (including
L4) are indicated by a red dashed line along this curve.
 
To conclude this section, let us summarize the main dynamical behaviours
observed in the logarithmic model. For the stable configurations, we find
that small perturbations remain bounded and the fields remain relatively close
to their initial values without any sign of significant growth or decay. For
the unstable configurations, we find the most common outcome to be
fragmentation into a small number of ``child" Q-balls which quickly propagate
away along the axis of symmetry. In some cases (such as L3), this process is
accompanied by the formation of Q-rings, while in other cases (such as L2), no
significant Q-ball or Q-ring remnants are formed. At present, we
have been unable to identify a simple criterion which can predict these changes
in behaviour. In general, the fragmentation of gauged Q-balls appears to be a
complicated non-linear process, with the only guarantee being the conservation
of four-momentum and charge.

\subsection{$V_6$ Model}

Here we consider the dynamical stability of gauged Q-balls in the polynomial
model \eqref{eqn:poly}.
For illustrative purposes, we select two configurations P1 and P2 whose
properties are listed in Table \ref{table:evos}. P1 represents an example of a
stable evolution for ${e}=0.17$ while P2 represents an unstable evolution
for ${e}=0.02$. For numerical purposes, we set $m=k=1$ and
${h}=0.2$ in all evolutions.

\begin{figure}
    \includegraphics[width=\columnwidth]{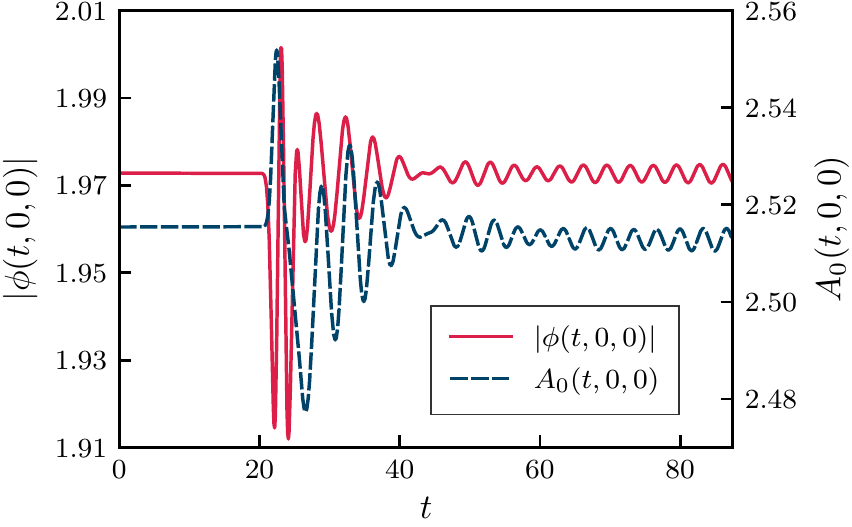}
    \caption{\label{fig:P1-osc} Oscillations of the scalar field modulus
    $|\phi|$ for configuration P1 subject to a Type I perturbation with
    parameters $c=0.1$ and $A=0.1$. Also shown are the oscillations in the
    maximal value of $A_0$ (right axis). The perturbative field $\chi$ implodes
    upon the Q-ball at $t\approx20$ and induces oscillations which quickly decay. By
    $t\approx60$, the perturbed Q-ball has returned very nearly to its original
    configuration.}
\end{figure}

First we consider the evolution of P1. This configuration lies on the shortest
curve of Fig.~\ref{fig:solnspace-poly} with ${e}=0.17$, which is near the
maximum allowable value of ${e}\approx0.182$ \cite{Loginov2020}. The evolution
of P1 is subject to a Type I perturbation with parameters $A=0.1$ and $c=0.1$ and
runs for 65000 base-grid time steps up a final time of $t=6400$.  The maximal
value of the scalar field modulus $|\phi|$ and the gauge field component $A_0$
for this evolution is shown in Fig.~\ref{fig:P1-osc}. The perturbative scalar
field hits the Q-ball at $t\approx20$ before exploding outward and exiting the
simulation domain.  After the collision, the Q-ball is left oscillating at the
origin around the stationary (unperturbed) configuration.  However, the
magnitude of this oscillation rapidly decays as the Q-ball quickly returns
close to the original configuration.  Similar behaviour to P1 is observed for
all other solutions tested on the ${e}=0.17$ branch depicted in
Fig.~\ref{fig:solnspace-poly}. We therefore conclude that P1 (as well as all
other solutions tested on this branch) is dynamically stable.

\begin{figure} [t]
\includegraphics[width=\columnwidth]{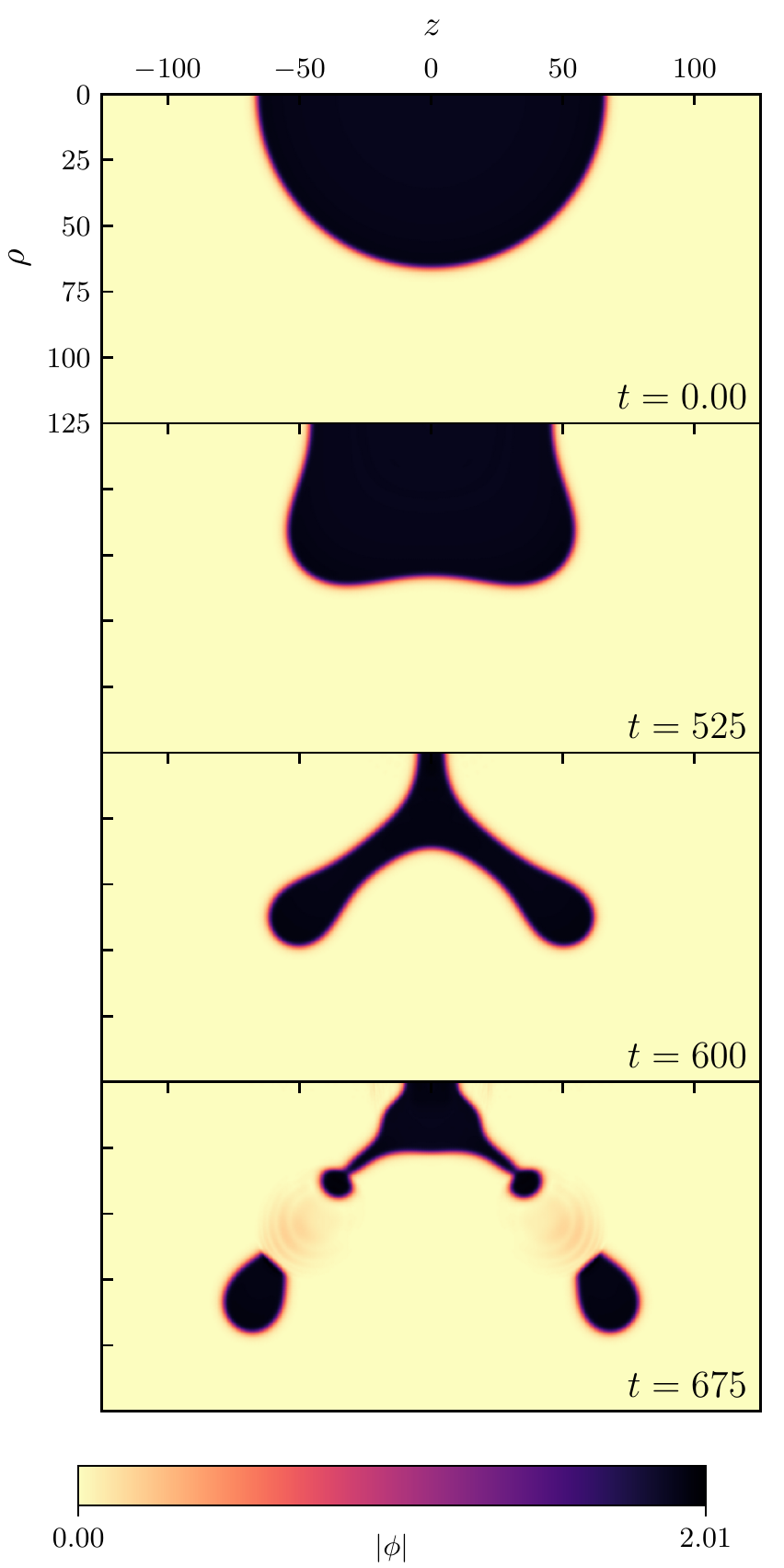}
  \caption{\label{fig:P2-evo} Evolution of the scalar field modulus $|\phi|$
  for configuration P2 subject to a Type 0 perturbation. As the evolution proceeds,
  the Q-ball splits into two Q-rings which propagate away from the axis of symmetry. The axis of symmetry is coincident with the top edge of each panel.  The simulation
  domain spans $\{\rho\,:\,0\le\rho\le 150\}$ and $\{z\,:\,-150\le z \le 150\}$.}
\end{figure}

Finally, we consider the evolution of P2. This configuration is distinctive in
that it occupies a much larger volume than any of the configurations previously
considered. In addition, the scalar field profile of P2 is relatively uniform
in the center of the Q-ball before dropping off rapidly to zero at a radial
distance $r\approx65$. In this sense it somewhat resembles a Q-ball of the
thin-wall type \cite{Heeck2021c}.
In Fig.~\ref{fig:P2-evo}, we show the evolution of P2 subject
to a Type 0 perturbation.  The distinctive flat-top shape of the configuration is
apparent in the first panel of the figure. 
By $t\approx525$ (second
panel), the original spherical shape of the configuration is lost as the field
content begins to concentrate away from the axis of symmetry. At late times,
these off-axis concentrations separate into two distinct Q-rings while a relic
region of Q-matter remains near the origin.

\begin{figure} [t]
\includegraphics[width=\columnwidth]{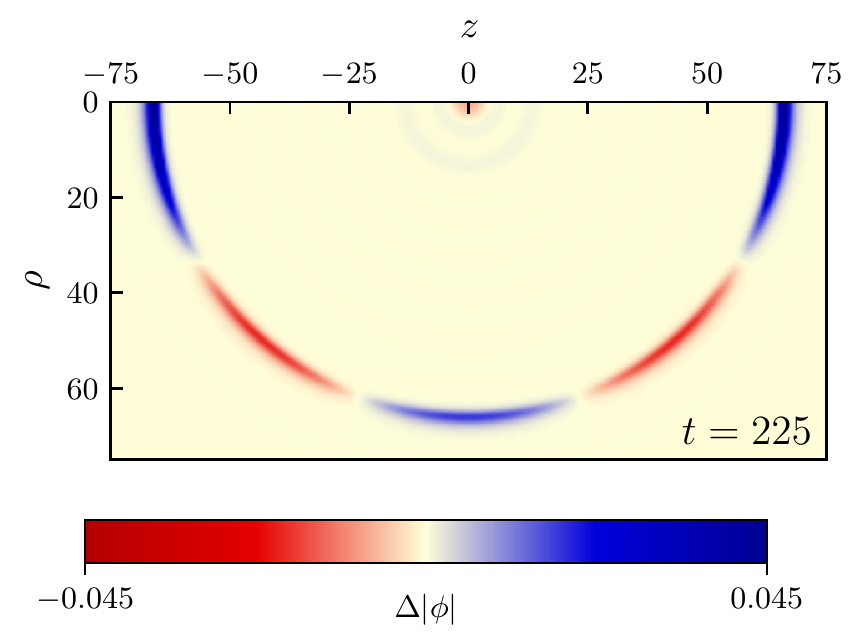}
  \caption{\label{fig:P2-diff}
  Plot of the difference in the scalar field modulus
  $\Delta|\phi|=|\phi(t=0,\rho,z)|-|\phi(t=225,\rho,z)|$
  for configuration P2 subject to a Type 0 perturbation. Here, the growth of $|\phi|$ occurs predominantly near the edge of the Q-ball and resembles the pattern of the $Y_{4,0}$ spherical harmonic. The corresponding plot for the full evolution of $|\phi|$ is given in Fig.~\ref{fig:P2-evo}.}
\end{figure}

In Fig.~\ref{fig:P2-diff}, we plot the growth of the scalar field modulus
$|\phi|$ for configuration P2. Here, the difference
$\Delta|\phi|=|\phi(t=0,\rho,z)|-|\phi(t=225,\rho,z)|$ illustrates the growth
of the solution between the initial time and at a point midway through the
evolution (but before the dynamics have entered the non-linear regime). It
is clear from the figure that the growth occurs predominantly near the edge of
the Q-ball and resembles the pattern of the $Y_{4,0}$ spherical harmonic.
This pattern becomes apparent in the evolution by $t\approx 100$ and grows
exponentially in amplitude until the Q-ball begins to break apart starting at
$t\approx 500$. As mentioned previously, it is well-known that the decay of
unstable non-gauged Q-balls is always mediated by a spherically-symmetric mode
\cite{Smolyakov2018}.  However, it remains an open question in the literature
as to whether gauged Q-balls can be destroyed by the growth of non-spherical
modes. Here we have found an example of an unstable gauged Q-ball where the
growth of the dominant unstable mode appears to be non-spherical.
Remarkably, this occurs even for a small gauge coupling value of
${e}=0.02$. This result is suggestive (though not conclusive) that the
destruction of gauged Q-balls can be mediated by non-spherical modes, in
contrast to their non-gauged counterparts. However, we emphasize that we have
not made any attempt to perform a full stability analysis in this work.

In Fig.~\ref{fig:solnspace-poly-stab}, we plot the location of P2 on the
${e}=0.02$ curve in the solution space.  The curve can be broken down
into several distinct branches: an upper unstable branch (I), a stable
branch (II), and a lower unstable branch (III) which contains P2.  The
branch (III) is characterized by solutions which are dominated by a
large, nearly-homogeneous central region and have thin surface
boundaries, similar to P2. The radial extent of these solutions
increases with ${\omega}$ along branch (III).
In most cases, the gauged Q-balls on this branch are found to decay slowly into
smaller gauged Q-balls or Q-rings, in contrast to the unstable branch (I) where
the instability quickly manifests via complete dispersal of the fields to
infinity (similar to L2 in the logarithmic model). However, for some
configurations along branch (III) which are close to the transition point with
branch (II), we also observe the development of large oscillations in the
Q-ball interior which significantly disrupt the shape of the configuration but do not cause
the Q-ball to immediately break apart.  These oscillations are accompanied by
the radiation of charge toward infinity. Since these solutions lose their
resemblance to the initial configuration, we also classify them as unstable.

Before concluding, let us comment on the validity of the classical stability
criterion \eqref{eqn:VK-criteria} for ${e}=0.02$.  Since the gauge
coupling is very small for this case, one might expect that the regions of
stability should be well-predicted by the sign of $dQ/d\omega$.  Indeed, we
find this to be the case. The unstable branches (I) and (III) in
Fig.~\ref{fig:solnspace-poly-stab} approximately correspond with
{$(\omega/Q)\,dQ/d\omega>0$ while solutions on the stable branch (II)
approximately correspond with $(\omega/Q)\,dQ/d\omega<0$. In this sense the gauged
Q-balls with ${e}=0.02$ closely resemble their non-gauged counterparts.  In
contrast, we have found the entire space of solutions for ${e}=0.17$ to be
stable despite the fact that one can find both $(\omega/Q)\,dQ/d\omega>0$ and
$(\omega/Q)\,dQ/d\omega<0$ for these solutions. This supports the finding that the sign of
the criterion \eqref{eqn:VK-criteria} provides little information on the
classical stability of gauged Q-balls when the magnitude of the gauge coupling
is appreciable \cite{Panin2017}.

\begin{figure}
    \includegraphics[width=\columnwidth]{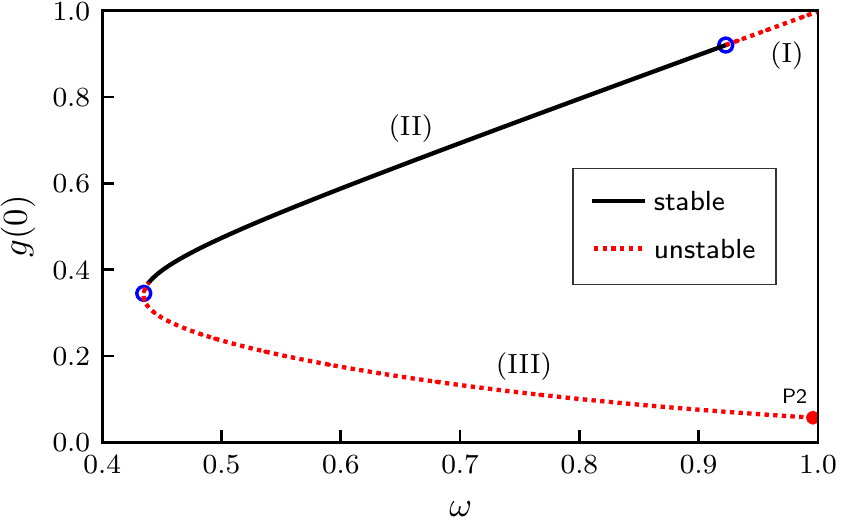}
    \caption{\label{fig:solnspace-poly-stab} Regions of stability and
    instability for gauged Q-balls in the $V_6$ model with ${e}$=0.02.
    The point P2 corresponds with the configuration listed in Table
    \ref{table:evos}. The dotted red line represents unstable regions (I) and
    (III) while the solid black line (II) corresponds with stable
    configurations. The blue circles indicate the transition points between
    stability and instability as predicted by the classical stability criterion
    for non-gauged Q-balls \eqref{eqn:VK-criteria}.}
\end{figure}

\section{\label{sec:conclusion} Conclusion}

We have performed fully non-linear numerical evolutions of $U(1)$ gauged
Q-balls in axisymmetry to investigate their stability and dynamics. We assessed
this stability in two ways: by perturbing the configurations using the inherent
truncation error of the numerical grid, and by introducing an auxiliary
massless real scalar field which acts as a perturbing agent designed to
explicitly excite any underlying unstable modes. Our simulations suggest that
both stable and unstable gauged Q-ball configurations can exist in both the
logarithmic and polynomial models. For those solutions which are classified as
stable, we observe no evidence of growing modes on the timescales of our
evolutions.  These solutions respond to perturbations by oscillating
continuously or weakly radiating before returning to the original
configuration.  On the other hand, the decay of unstable configurations can
manifest in several different ways, such as total dispersal of the solution,
fragmentation into smaller gauged Q-balls, or via the formation of
non-spherical ring-like structures which we call ``gauged Q-rings".  These
structures appear to be similar in appearance and behaviour to the rings
observed in previous studies of non-gauged Q-ball dynamics \cite{Battye2000}.
Additionally, for some solutions governed by
the logarithmic potential and which attain
large field values, we observe that the configurations respond to
perturbations by growing without bound. This is similar to behaviour observed
in other Q-ball models that permit a negative energy density and is
interpreted as a
consequence of the scalar potential being unbounded from below. For the
polynomial potential, we have also investigated the dynamical behaviour of
gauged Q-balls when the gauge coupling is small. In this case, we find that the
regions of stability and instability are well-described by the
stability criterion \eqref{eqn:VK-criteria}.

One expected result from our study is that those configurations which were
known to be unstable with respect to spherically-symmetric perturbations
\cite{Panin2017} are also unstable under axisymmetric ones. However, our
results indicate that axisymmetric perturbations can lead to new regions of
instability in the solution space. Furthermore, we have found that some
unstable gauged Q-ball configurations can be destroyed by the growth of modes
which appear to be non-spherical. These results suggest that the decay of some
gauged Q-ball configurations may be mediated by non-spherical modes, in
contrast to non-gauged Q-balls. 

While we have presented numerical evidence that gauged Q-balls can be
classically stable with respect to axisymmetric perturbations, it is possible
that more general perturbations may eventually destroy any gauged Q-ball.
Addressing this issue might be accomplished with a fully three-dimensional
code. Future work may also focus on trying to explicitly solve for the
non-spherical gauged Q-ring configurations and numerically evolving them in
order to assess their stability. Another interesting question relates to the
interaction of two stable gauged Q-balls, which will be the subject of a future
paper.

Lastly, we would like to emphasize that our results have addressed only the
classical stability of gauged Q-balls (i.e.,~stability of the solutions with
respect to small perturbations of the fields). For a complete picture of Q-ball
behaviour, one may also wish to consider quantum effects. For example, a Q-ball
may decay through collective tunnelling or by surface evaporation when coupled
to additional fields \cite{Tranberg2014,Levkov2017,Cohen1986,Hong2017}. It is
interesting to ask whether the stability of gauged Q-balls is maintained once
these effects are considered, though such a question is beyond the scope of the
present work.

\acknowledgements

We thank G.~Reid for providing valuable advice and comments. 
This work was supported by the Natural Sciences and Engineering Research
Council of Canada (MWC, MPK), the Walter C.~Sumner
Foundation (MPK) and the Province of British Columbia (MPK). 
Numerical simulations were performed on the Graham and Cedar clusters of the 
Digital Research Alliance of Canada.\\

\appendix

\section{Evolution Equations in Axisymmetry} \label{apx:evo-eqns}

Here we present the full set of evolution equations for our model, the
boundary conditions used, and the regularity conditions imposed on the
axis of symmetry. 

The evolution equations for the scalar and gauge
fields can be expressed in cylindrical coordinates as
\begingroup
\allowdisplaybreaks
\begin{widetext}
\begin{align}
  \partial_t^2\phi_1=&\;\frac{1}{\rho}\partial_\rho\phi_1+\partial_\rho^2\phi_1+\partial_z^2\phi_1+2e\left(-A_t\partial_t\phi_2+A_\rho\partial_\rho\phi_2+A_z\partial_z\phi_2 \right) -e^2\phi_1\left(-A_t^2+A_\rho^2+A_z^2+\frac{1}{\rho^2}A_\varphi^2\right)-\frac{1}{2}\partial_{\phi_1}V(\phi_1,\phi_2), \label{eqn:phi1res}\\
    \partial_t^2\phi_2=&\;\frac{1}{\rho}\partial_\rho\phi_2+\partial_\rho^2\phi_2+\partial_z^2\phi_2-2e\left(-A_t\partial_t\phi_1+A_\rho\partial_\rho\phi_1+A_z\partial_z\phi_1 \right)-e^2\phi_2\left(-A_t^2+A_\rho^2+A_z^2+\frac{1}{\rho^2}A_\varphi^2\right)-\frac{1}{2}\partial_{\phi_2}V(\phi_1,\phi_2),\\
  \partial_t^2 A_t=&\;\frac{1}{\rho}\partial_\rho A_t+\partial_\rho^2 A_t+\partial_z^2 A_t+2e \left( \phi_1 \partial_t \phi_2 - \phi_2 \partial_t \phi_1 \right) - 2e^2 \left( \phi_1^2+\phi_2^2 \right)A_t,\\
  \partial_t^2 A_\rho=&\;\frac{1}{\rho}\partial_\rho A_\rho+\partial_\rho^2 A_\rho+\partial_z^2 A_\rho-\frac{A_\rho}{\rho^2}+2e \left( \phi_1 \partial_\rho \phi_2 - \phi_2 \partial_\rho \phi_1 \right) - 2e^2 \left( \phi_1^2+\phi_2^2 \right)A_\rho,\\
  \partial_t^2 A_\varphi=&\;-\frac{1}{\rho}\partial_\rho A_\varphi+\partial_\rho^2 A_\varphi+\partial_z^2 A_\varphi-2e^2 \left( \phi_1^2+\phi_2^2 \right)A_\varphi,\\
  \partial_t^2 A_z=&\;\frac{1}{\rho}\partial_\rho A_z+\partial_\rho^2 A_z+\partial_z^2 A_z+2e \left( \phi_1 \partial_z \phi_2 - \phi_2 \partial_z \phi_1 \right) - 2e^2 \left( \phi_1^2+\phi_2^2 \right)A_z, \label{eqn:Azres}
\end{align}
\end{widetext}
\endgroup
\noindent
where the subscripts $\{t,\rho,\varphi,z\}$ correspond to the spacetime 
coordinates and the subscripts $\{1,2\}$ correspond
to the real and imaginary parts, respectively, of the scalar field $\phi$. For numerical purposes,
we find it convenient to define new evolutionary variables
\begin{equation}
  \begin{alignedat}{4} \label{eqn:first-order}
     \Pi_1 &=\partial_t\phi_1, &\qquad \Pi_2 &=\partial_t\phi_2,\\
     G_t &=\partial_t A_t,     &\qquad G_\rho &=\partial_t A_\rho,\\
     G_\varphi &=\partial_t A_\varphi, &\qquad G_z &=\partial_t A_z,
  \end{alignedat}
\end{equation}
which are first-order in time. 

To complete our specification of the problem, we
must impose appropriate boundary conditions along the axis of symmetry and at the outer domain boundaries $\rho=\rho_\text{max}$, $z=z_\text{min}$, and
$z=z_\text{max}$. For the outer boundaries, we apply outgoing boundary conditions of the form
\begin{equation} \label{eqn:sommerfeld}
  \sqrt{\rho^2+z^2}\;\partial_t f + f + \rho\,\partial_\rho f + z\,\partial_z f = 0,
\end{equation}
where $f=f(t,\rho,z)$ represents each of the evolved variables in
\eqref{eqn:phi1res}--\eqref{eqn:first-order}. Strictly speaking, the condition
\eqref{eqn:sommerfeld} assumes the field $f$ is massless and purely radially
outgoing at the domain boundary. For non-spherical pulses or massive fields,
the approximation can break down and lead to unphysical reflections. However,
these partial reflections can be mitigated by taking the outer boundaries 
sufficiently far away and by stopping the evolution when appreciable field content
reaches the boundaries. With these considerations, we find condition
\eqref{eqn:sommerfeld} to be sufficient for the purposes of this work.

Along the axis of symmetry, we find it useful to define the regularized variables
\begin{equation}
  \begin{alignedat}{4}
    \tilde{G}_\rho &=\frac{G_\rho}{\rho}, &\qquad \tilde{G}_\varphi &=\frac{G_\varphi}{\rho^2}.\\
  \end{alignedat}
\end{equation}
With this definition, we identify $G_t$, $\tilde{G}_\rho$, $\tilde{G}_\varphi$,
$G_z$, $\Pi_1$, and $\Pi_2$ as having even character as $\rho\rightarrow0$. We
can therefore demand that the radial derivative of these fields should vanish
in this limit, yielding appropriate regularity (boundary) conditions along the 
axis of symmetry.

\section{Code Validation}
\label{apx:numerics}

We have performed a number of tests in order to assess the validity of our
numerical code. For all calculations presented in this appendix, we evolve
generic Gaussian-like initial data which is smooth everywhere. The fields
$\phi_1$, $\phi_2$, $\Pi_1$, $\Pi_2$, $A_t$, $\tilde{A}_\rho$, and $A_z$ are chosen
to be arbitrary overlapping profiles of the form
\begin{equation} \label{eqn:generic-gauss}
  f(\rho,z)=A\exp\left[-\left(\frac{\sqrt{(\rho-\rho_0)^2+(z-z_0)^2}}{\delta}\right)^2\right],
\end{equation}
where $A$, $\delta$, $\rho_0$ and $z_0$ are real parameters which can be
different for each field. The fields $G_t$, $\tilde{G}_\rho$, and $G_z$ are
then found using a successive over-relaxation scheme \cite{Press2007} in order to
approximately satisfy the constraints at the initial time.

\begin{figure}
\includegraphics[width=\columnwidth]{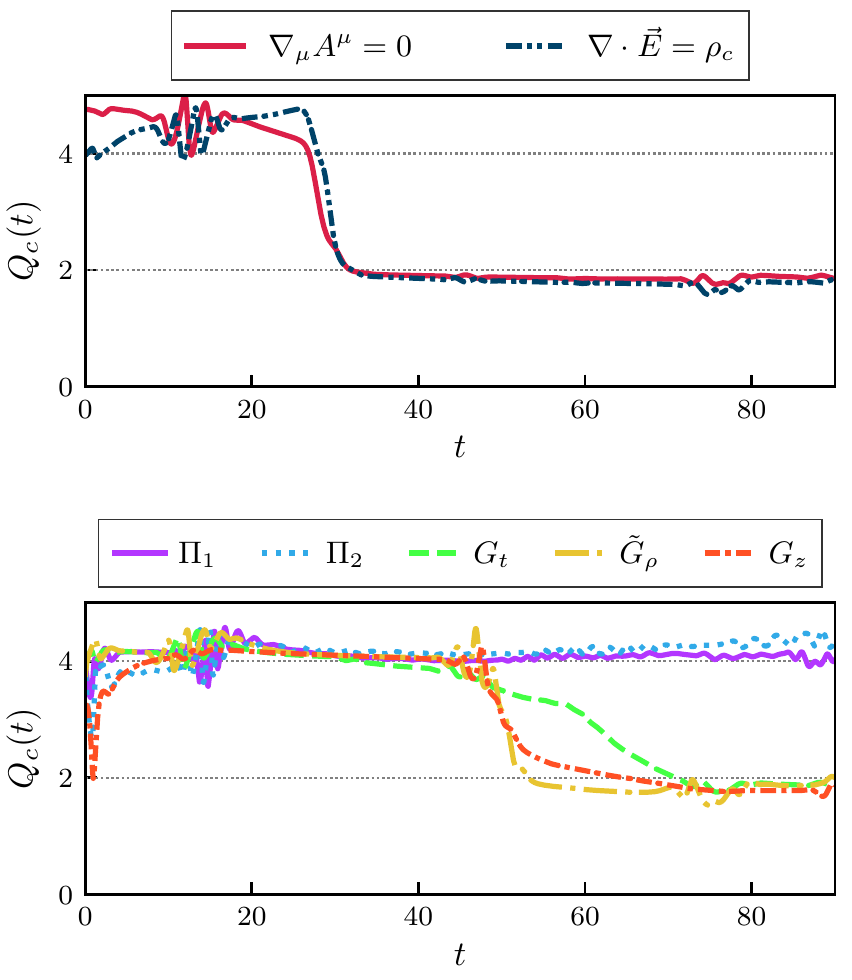}
  \caption{\label{fig:conv}Convergence factors $Q_c(t)$ for the constraint
  equations (top) and several first-order evolved variables $\Pi_1$, $\Pi_2$,
  $G_t$, $\tilde{G}_\rho$, and $G_z$ (bottom). Here, $Q_c(t)$ is computed using a
  three-level convergence test at resolutions $257\times 513$, $513\times 1025$
  and $1025\times 2049$. In each case, the quantities are found to be
  convergent at approximately second-order until the fields hit the boundaries,
  at which point first-order convergence is observed. For the data shown here,
  the potential \eqref{eqn:poly} is used with parameters $e=0.25$, $h=0.2$,
  and $m=k=1.0$.}
\end{figure}

\begin{figure}
\includegraphics[width=\columnwidth]{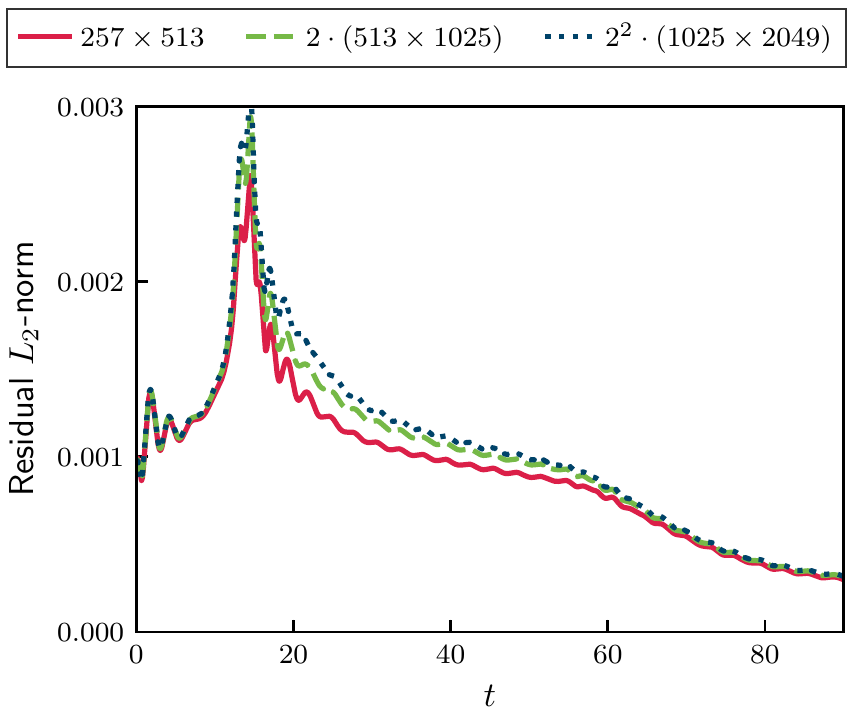}
  \caption{\label{fig:IRE} Residual $L_2$-norm values for the evolution
  equation \eqref{eqn:phi1res} computed at several different grid resolutions.
  The $L_2$-norm has been scaled by factors of $2^n$ for increasing
  resolutions. Overlapping of the curves indicates the expected first-order
  convergence of the residuals.}
\end{figure}

As a primary test, we evolve this generic Gaussian-like data on a uniform grid
at several different grid resolutions in order to explicitly compute the rate
of convergence of our code.
Let us define the convergence factor $Q_c(t)$ as
\begin{equation}
  Q_c(t)=\frac{\|u^{4h}-u^{2h}\|}{\|u^{2h}-u^{h}\|},
\end{equation}
where $h$ represents the spacing between points on the numerical grid, $u^n$
represents the solution computed with grid spacing $n$, and $\|\cdot\|$ denotes
the $L_2$-norm. For a second-order-accurate finite-difference scheme, it can be
shown that $Q_c(t)\rightarrow4$ as $h\rightarrow 0$. In
Fig.~\ref{fig:conv}, we plot the convergence factor resulting from our test for
the constraint equations and several representative fields. The rate of
convergence is found to be approximately second-order (corresponding to
$Q_c(t)=4$) which is to be expected for our second-order Crank-Nicolson
finite-difference implementation.

As a secondary measure, we have performed an independent residual test
\cite{Choptuik2006} to verify that our discrete numerical solution is
converging to the true continuum solution of the underlying system
\eqref{eqn:phi1res}--\eqref{eqn:Azres}. For this test, we substitute the
numerical solution found via the second-order Crank-Nicolson discretization
into a separate first-order forward discretization of the evolution equations.
Results of this test are shown in Fig.~\ref{fig:IRE}. The residuals are found
to approximately overlap when rescaled by factors of $2^n$, indicating the
expected first-order convergence.  For brevity, only the \eqref{eqn:phi1res}
residual is presented here -- other residuals are found to be similar.

In solving the equations of motion, we have used a free evolution scheme
wherein a solution to the evolution equations is expected to solve the
constraint equations at the initial time \cite{Choptuik1991}.  However, it is
possible for constraint violation to grow during the course of the evolution,
indicating lack of convergence.  The degree to which the constraints are
violated is therefore a relative measure of the error in the numerical
solution. In all of our simulations, we monitor the $L_2$-norm of the
constraint residuals to ensure that they do not grow significantly over the
timescales explored. For the results reported here, the $L_2$-norm of the
constraint residuals remains within $O(10^{-4})$. In addition, we also
monitor the integrated total energy $E$ and the charge $Q$ during the course of
the evolution to confirm that these quantities remain approximately conserved
to within $O(1\%)$.


\bibliographystyle{apsrev4-2}
\bibliography{refs}  

\end{document}